\documentclass[
 reprint,
 superscriptaddress,
 amsmath,amssymb,
 aps,
 pra
]{revtex4-2}

\usepackage{graphicx}

\begin{document}

\title{Designing metasurface optical interfaces for solid-state qubits using many-body adjoint shape optimization}

\author{Amelia R. Klein}
\affiliation{Department of Electrical and Systems Engineering, University of Pennsylvania, Philadelphia PA 19104, USA}

\author{Nader Engheta}
\affiliation{Department of Electrical and Systems Engineering, University of Pennsylvania, Philadelphia PA 19104, USA}

\author{Lee C. Bassett}
\email{lbassett@seas.upenn.edu}
\affiliation{Department of Electrical and Systems Engineering, University of Pennsylvania, Philadelphia PA 19104, USA}

\begin{abstract}
We present a general strategy for the inverse design of metasurfaces composed of elementary shapes.
We use it to design a structure that collects and collimates light from nitrogen-vacancy centers in diamond.
Such metasurfaces constitute scalable optical interfaces for solid-state qubits, enabling efficient photon coupling into optical fibers and eliminating free-space collection optics.
The many-body shape optimization strategy is a practical alternative to topology optimization that explicitly enforces material and fabrication constraints throughout the optimization, while still achieving high performance.
The metasurface is easily adaptable to other solid-state qubits, and the optimization method is broadly applicable to fabrication-constrained photonic design problems.
\end{abstract}

\maketitle

\section{Introduction}

Solid-state quantum defects\,---\,exemplified by the nitrogen-vacancy (NV) center in diamond---have emerged at the forefront of quantum technologies in recent years, seeing applications in core research areas of quantum computing \cite{Awschalom2018QuantumSpins, Abobeih2022}, communication \cite{Bradley2019AMinute, Pompili2021RealizationQubits, Bersin2024TelecomMemory}, and sensing \cite{Zhou2020QuantumSystems, Burgler2023All-opticalDiamond, Li2022SARS-CoV-2Diamond}. 
Measurements are performed optically, with the defect's spin state being encoded in its single-photon photoluminescence. 
As a result, efficient photon capture is essential, with readout fidelity, sensitivity, and operational speed being limited by optical losses. 
One fundamental source of optical loss is diamond's high refractive index at optical wavelengths (around 2.4), which causes the majority of emitted photons to be trapped at the flat air-diamond interface due to total internal reflection. 
Past works have addressed this challenge by embedding NV centers in nanophotonic structures such as waveguides, nanopillars, and photonic crystal cavities\cite{Mouradian2015ScalableCircuit, Neu2014PhotonicDiamond, Barclay2009HybridNV-centers, Chakravarthi2023HybridStamp-Transfer, Zheng2023PyramidalNanodiamond}; such structures in the near-field overcome total internal reflection and increase the photon emission rate through Purcell enhancement. 
However, near-field structures also situate NV centers near diamond surfaces, which conversely degrade their crucial spin and optical features \cite{Ofori-Okai2012SpinDiamond, Rosskopf2014InvestigationDiamond, Chou2023AbSpins}.

For applications such as quantum memories in which maximizing the spin coherence time is paramount, it is necessary to use NV centers embedded deeper in the bulk diamond and to only modify the far-field propagation of emitted photons. 
The prototypical structure used in this scenario is the solid immersion lens (SIL), which consists of a hemisphere centered around a defect, such that any emitted photon will be incident normal to the air-diamond interface and not experience total internal reflection. 
These hemispheres are typically etched into the diamond \cite{Castelletto2011Diamond-basedLight} but can also be created with additive fabrication methods \cite{Cheng2023AdditiveCenters}. 
However, SILs require a difficult and time-consuming volumetric fabrication process, and they do not adjust the propagation direction of emitted light, so high numerical aperture (NA) free-space microscope objectives are still required to collimate the diverging beam. 

In contrast, flat optics etched on the diamond surface can both transmit \textit{and} redirect incident photons \cite{Li2015EfficientGrating, Huang2019AEmitters}. 
Flat optics are easier and cheaper to fabricate in diamond than volumetric structures like SILs, making them an attractive option for scaled-up quantum computation or quantum communication systems. 
In addition, by simultaneously collimating the emitted photons, a flat optic can circumvent the need for expensive, bulky, free-space optics and ultimately allow the photons to be collected by an optical fiber in a compact, robust, inexpensive device.

In this work, we use inverse design methods to design a monolithic metasurface in diamond tailored to maximize the directional collection of photons emitted from an NV center. 
In contrast to other flat optics that are designed by local approximations, inverse design techniques capture the complete optical response of a metasurface geometry through full-wave simulations and target a figure of merit that represents explicitly the experimentally-relevant metric\,---\,in this case, the total photon collection efficiency from an NV center, averaged over multiple optical dipole orientations as well as the broadband photoluminescence spectrum.

\section{Many-Body Shape Optimization}
\subsection{Background}

\begin{figure*}
    \centering
    \includegraphics{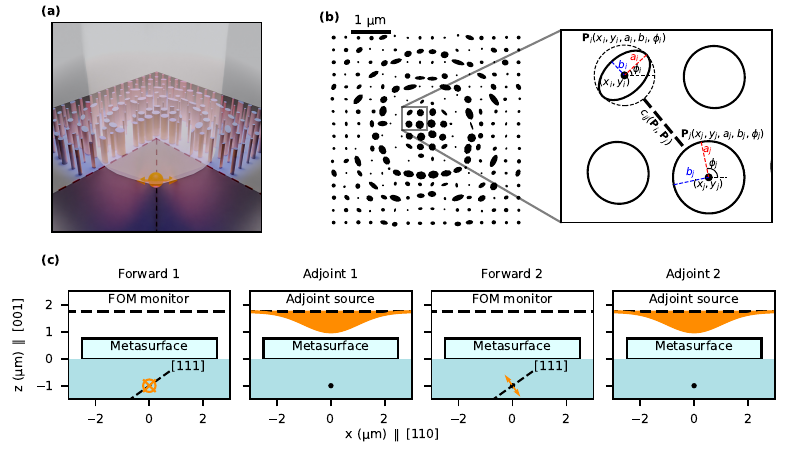}
    \caption{\textbf{(a)} To-scale render of optimized collimating metasurface. \textbf{(b)} Top-down view of nanopillar array. Inset: Zoom-in on one section, showing the parameterization of each nanopillar. Since the center points of each ellipse are free parameters, the structure is not constrained to a fixed grid like most conventional metasurfaces, but constraint functions enforce a minimum distance between adjacent elements for fabrication considerations. \textbf{(c)} Diagram of simulations run during each step of adjoint optimization. Two forward and adjoint simulation pairs correspond to two different dipole orientations (orange arrows) that are mutually orthogonal to each other and to the [111] NV axis. This set of simulations determines a figure of merit and its gradient, which is fed into a gradient-descent optimization algorithm.}
    \label{fig:concept}
\end{figure*}

We utilize an inverse design method based on the adjoint method, which has been utilized extensively in the photonics field in recent years for applications varying in complexity from simple components to arbitrary equation solvers \cite{Molesky2018InverseNanophotonics, Estakhri2019, Cordaro2023SolvingMetagratings}. 
With just two physics simulations\,---\,a typical "forward" simulation of optical propagation and an "adjoint" simulation that approximately represents the desired outcome sent backwards\,---\,an arbitrary figure of merit and its gradient can be calculated with respect to any number of design parameters. 
This can then be wrapped into a gradient-descent optimization algorithm to arrive at an optimized structure. 

Adjoint-based optimizations are categorized based on their parameterization method as either topology optimization or shape optimization. 
Most adjoint-based methods in recent years have made use of topology optimization, which treats the relative permittivity $\epsilon$ at each point in the optimization region as a free continuous parameter in order to explore a maximally broad parameter space. 
Typically, binarization and smoothing filters are applied gradually over the course of the optimization to ensure a realistic final structure made of discrete materials (\textit{e.g.,} using only two values of $\epsilon$) with smooth features \cite{Bendse1999MaterialOptimization}. 
Previous authors have used topology optimization to design optical metasurfaces based on patterned thin films \cite{Chung2020High-NADesign, Lin2019TopologyMetasurfaces, Wang2019RobustMetasurfaces} and nanophotonic structures using diamond membranes \cite{Dory2019Inverse-designedPhotonics}. 
Topology optimization has also been utilized to design near-field antenna-like structures to extract photons from shallow NV centers \cite{Chakravarthi2020Inverse-designedQubits, Wambold2020}.

In contrast, shape optimization considers a fixed number of objects, whose boundaries are parameterized according to their size, position, orientation, \textit{etc,}. 
Shape optimizations have been performed on individual objects to design structures such as Y-splitters \cite{Lalau-Keraly2013} and to generate unit cell libraries for larger metasurfaces \cite{Whiting2020Meta-atomMethod}. 
Other works have performed shape optimizations on periodically ordered arrays of objects \cite{Zhan2018InverseSpheres, Mansouree2021Large-ScaleOptimization} and for disordered photonic crystals \cite{Minkov2020InverseDifferentiation, Granchi2023Q-FactorTheory}.
Additionally, we became aware of similar works performing inverse design on metasurfaces using freely-moving pillars while this manuscript was under review \cite{Zhou2024Inverse-designedParameters, Zhou2024Large-AreaMetasurfaces}.

\subsection{Implementation}

In this work, we parameterize the metasurface as an array of elliptical diamond nanopillars and perform a many-body shape optimization (MBSO) over the entire set. 
Compared to unit-cell-based designs, we explore a broader parameter space by allowing the \textit{positions} of the nanopillars, in addition to their sizes, ellipticity, and orientation, to be free parameters in the optimization. 
Each nanopillar $P_i$ is defined by a fixed height and an elliptical cross-section described by five optimization parameters $P_i(x_i, y_i, a_i, b_i, \phi_i)$: two center coordinates, two axial lengths, and one rotation angle. 
Nanopillar heights are fixed uniformly at a predefined value such that the final surface could be fabricated with a single etch. 
Figures~\ref{fig:concept}a and \ref{fig:concept}b show an optimized metasurface along with these shape-optimization parameters.

The fabrication of diamond nanopillar arrays is well established. 
They are used, for example, in quantum sensing applications in which NV centers are embedded directly in the nanopillars \cite{Zou2008FabricationArrays, Hausmann2011Single-colorNanostructures, Losero2023NeuronalApplications}. Previous work has also shown how diamond nanopillar arrays can function as a metalens to collect light from deep NV centers \cite{Huang2019AEmitters}, motivating the present work. 
The device efficiency and NA of the metalens in Ref.~\cite{Huang2019AEmitters} was limited by the classical design strategy, which treated each pillar as a localized phase-shifting element.

The explicit formulation of the metasurface as an array of nanopillars ensures a fabricable design at each iteration of the optimization process. 
Fabrication constraints are enforced through lower bounds on the axial lengths to ensure a minimum pillar size as well as differentiable inequality constraint functions between nearby pillar pairs $(P_i, P_j)$ to ensure a minimum separation, $d$:
\begin{equation}\label{eq:constraints}
    c_{ij}(P_i, P_j) = \sqrt{(x_i-x_j)^2 + (y_i - y_j)^2} - r_i - r_j \geq d\\
\end{equation}
Since there is no analytical expression for the distance between ellipses, the elliptical nanopillars are instead conservatively approximated as circles with $r_i= \max(a_i, b_i)$ for the constraint calculations. 
These constraint functions are applied between adjacent pillars based on their initial configuration on a rectangular grid, including vertically, horizontally, and diagonally adjacent pairs.
While pillars are allowed to move freely over the course of the optimization, constraints based only on their initial positions proved sufficient for enforcing a minimum gap size in the optimizations shown in this work.
To perform the optimizations, we use these constraint functions along with adjoint-calculated gradients in an open-source implementation of the Method of Moving Asymptotes algorithm \cite{Svanberg2002AApproximations, NLopt}.

\subsection{Figure of Merit}

We construct the figure of merit for adjoint optimization to maximize the percentage of emitted photons that escape the diamond surface into a desired collection angle. 
We consider the full NV emission spectrum, which at room temperature is dominated by a wide phonon sideband spanning $\approx$650-750~nm.
At a collection plane above the NV surface, the simulated electric and magnetic fields are expressed as a function of propagation directions by taking a spatial Fourier transform.
These fields are filtered according to the function $u(k_x, k_y)$ to include only the fields within the acceptance angle of a large, multi-mode fiber with a numerical aperture of $\mathrm{NA_{target}}$ placed above the metasurface:
\begin{equation}
u(k_x, k_y) = \begin{cases}
            1 & \sqrt{k_x^2 + k_y^2}/k_0 \leq \mathrm{NA_{target}}\\
            0 & \mathrm{else}
            \end{cases}
\end{equation}
where $k_x$ and $k_y$ are the $x$ and $y$ wavevector components and $k_0 = \frac{2\pi}{\lambda_0}$. 
For a given dipole orientation \textit{i}, the filtered electric fields are described by applying this filter on the Fourier-transformed fields and then taking an inverse Fourier transform to return to real space:
\begin{equation}
\mathbf{E}_{\mathrm{filt}}^i(\mathbf{r}) = \mathcal{F}^{-1}\Big[{u(k_x, k_y)\mathcal{F}[\mathbf{E}^i(\mathbf{r})]}\Big]
\end{equation}
where $\mathcal{F}$ and $\mathcal{F}^{-1}$ describe the spatial Fourier transform and its inverse, and the filtered magnetic fields $\mathbf{H}_{\mathrm{filt}}^i(\mathbf{r})$ are calculated analogously.
These real-space filtered fields are needed to calculate the source for the adjoint simulations as described in Supplement 1 and are also integrated over the collection plane to obtain the directional transmission into $\mathrm{NA_{target}}$:
\begin{equation}
f_i(\lambda) = \frac{1}{P_\mathrm{dipole}}\iint\frac{1}{2}\operatorname{Re}\big(\mathbf{E}_{\mathrm{filt}}^i\times\mathbf{H}_{\mathrm{filt}}^{i*})\cdot \boldsymbol{\hat{\textbf{z}}} dA
\end{equation}
where $P_\mathrm{dipole}$ is the total power injected into the simulation.
The total figure of merit (FOM) is subsequently constructed as follows:
\begin{equation}\label{eq:spectralavg}
\text{FOM} = \frac{1}{2}\sum_{i=1,2}\int f_i(\lambda)\eta_\mathrm{NV}(\lambda)d\lambda
\end{equation}
where $\eta_\mathrm{NV}(\lambda)$ is the normalized room-temperature NV emission spectrum, and $i=1,2$ represent two dipole orientations that are mutually orthogonal to each other and to the NV-axis.

We consider a diamond with a top surface aligned with a (100) crystal plane, which is typical of synthetic diamond plates used in NV-center experiments. 
The NV center itself is oriented along a $\langle111\rangle$ axis, and its emission pattern results from an incoherent sum of two optical dipoles orthogonal to each other and to the NV axis, as illustrated in Fig. \ref{fig:concept}c. 
Less commonly, diamond can be grown \cite{Tallaire2014HighSubstrates} or cleaved \cite{Parks2018FabricationCleaving} to create a (111) top surface. 
The (111)-oriented diamond hosts NV centers whose optical dipoles are oriented parallel to the surface and subsequently achieves higher efficiency. 
The optimization process can be easily performed under the assumption of (111) diamond; see Supplement 1 for a comparison of the final results.

The source for the adjoint simulation is derived from the figure of merit, and the gradients are calculated by integrating the forward and adjoint fields over the nanopillar surfaces according Eqns. (S1-S5) in the Supplement; see also Ref.~\cite{Miller2012}. 
The simulations required to perform this optimization are depicted in Fig.~\ref{fig:concept}c.

In order to simultaneously optimize over a broad frequency range, we perform simulations in the time domain using Lumerical FDTD \cite{Lumerical}. 
The optimization procedure is implemented using the LumOpt package \cite{Lalau-Keraly2013} as a base for additional custom Python code, which implements the geometry construction, figure of merit and gradient calculations, and optimization algorithm wrapper (See Code 1, Ref. \cite{Lumopt-mbso}).

\section{Results}
\subsection{Performance}

\begin{figure}
    \centering
    \includegraphics{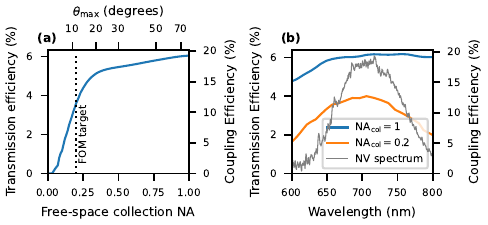}
    \caption{\textbf{(a)} Performance metric of final structure. Total photon transmission efficiency is calculated as a function of the numerical aperture of an ideal collection optic placed above the metasurface. The dashed line indicates the target used in the optimization. The right y-axis normalizes the transmission by the solid angle of the directional NV emission incident on the surface. \textbf{(b)} Performance is plotted as a function of wavelength for directional transmission efficiency (within $\mathrm{NA_{col}}$ = 0.2) and maximum transmission (within $\mathrm{NA_{col}}$ = 1). The gray line shows a normalized spectrum of a NV center measured at room temperature in \cite{Huang2019AEmitters} for weighting the spectral response to calculate the figure of merit.}
    \label{fig:characterization}
\end{figure}

We performed a many-body shape optimization to design a collimating collection metasurface for a NV center. 
We targeted a free-space collection numerical aperture of 0.2, roughly corresponding to a typical multi-mode fiber that could be placed above a metasurface. 
The design targeted an NV center centered 1~{\textmu}m beneath the metasurface. 
The height of the nanopillars composing the metasurface was fixed to 750 nm, and the total design area was 5~{\textmu}m $\times$ 5~{\textmu}m. 
We enforced a minimum pillar diameter and minimum gap size $d$ of 50 nm, using the formulation in Eqn. \ref{eq:constraints} for the latter. 
The resulting structure is depicted in Fig. \ref{fig:concept}a-b, and its performance is characterized in Fig.~\ref{fig:characterization}.

In order to benchmark device performance, we calculate the percentage of total photons emitted by an NV center that are collected into the desired free-space NA.
This total dipole transmission efficiency represents the most important performance metric, since it determines the photon count rate and the spin-readout signal-to-noise ratio.
It can also be easily compared across different configurations of diamond surface orientation, NV depth, and metasurface size.
For the simulated geometry, only 31.4\% of emitted photons are incident on the surface, and hence the right axes in Fig. \ref{fig:characterization} are normalized by this factor to show the metasurface coupling efficiency. 
The optimized structure exhibits a directional transmission efficiency (within NA$_\mathrm{target}$ = 0.2 and collectible by a multi-mode fiber) of 3.61\%, corresponding to a normalized coupling efficiency of 11.5\%. 
The maximum transmission efficiency (within NA = 1 and requiring a high-NA microscope objective) is 6.05\%, so the majority of outcoupled light is collimated into the targeted NA, despite NA$_\mathrm{target}$ representing only 4\% of possible propagation directions.

For comparison, we performed a topology optimization using the same objective function. 
We obtained a directional transmission efficiency of 3.80\% and a maximum transmission efficiency of 8.97\%. 
The topology-optimized structure has similar performance to the shape-optimized metasurface for the target NA, however it does transmit more of the light in total.
While the SIL's maximum transmission efficiency (24.1\%) is much higher, both the topology- and shape-optimized structures offer a nearly fourfold enhancement over the SIL's directional transmission efficiency of 1.03\%.
The performance of all three structures is described further in the supplementary information and plotted in Fig. S3.

Additionally, we demonstrate that our choice in allowing the position of pillars to be free parameters during the optimization results in a final design with a higher performance than one constrained to pillars fixed on a unit-cell grid.
We do this by performing an additional shape optimization in which the pillar positions are no longer free parameters, and obtain a directional transmission efficiency of 2.56\%, which is a 29\% decrease from the case in which pillars are allowed to move.
Both structures and their performance are plotted together in the supplementary information in Fig. S1.

In Figure~\ref{fig:colormaps}, we show the simulated cross-sectional and collection plane field profiles as an incoherent weighted average over the dipole orientations and wavelengths. 
The metasurface collimating effect is clearly visible in the real-space field profile of Fig. \ref{fig:colormaps}a and in the Fourier-space fields in Fig. \ref{fig:colormaps}c.

\begin{figure*}
    \centering
    \includegraphics{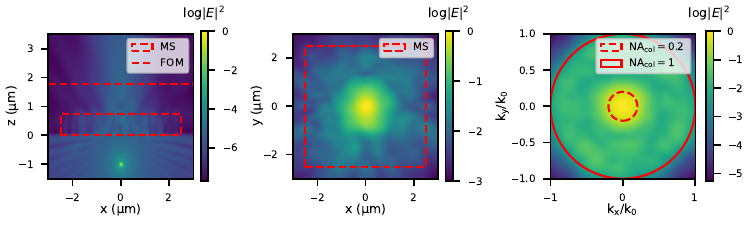}
    \caption{Electric field intensity profiles averaged incoherently across the broadband spectrum and orthogonal dipole orientations and normalized to the peak intensity in each plot. \textbf{(a)} xz cross-section of the electric field, plotted on a log scale. Dotted lines indicate the location of the metasurface (MS) and the collection plane used for figure of merit (FOM) calculations. \textbf{(b)} Electric fields at the plane of the FOM monitor. Dotted line indicates the extent of the metasurface beneath this plane. \textbf{(c)} Spatial Fourier transform of fields in \textbf{(b)}, plotted against normalized wavevector components $k_x/k_0$ and $k_y/k_0$. Dashed and solid lines indicate free-space numerical apertures of 0.2 and 1.}
    \label{fig:colormaps}
\end{figure*}

\subsection{Effect of Initial Conditions}\label{hyperparameters}

While the optimization process, by and large, determines the optimal parameters automatically, two key parameters are pre-selected and not variable during the optimization process. 
First, the nanopillar height was chosen to be a single, uniform value of 750 nm. 
In assuming a uniform height, we ensure that the final structure can be fabricated in a single etch step, but there is no technical reason why the pillar height could not be an additional optimization parameter.

The second pre-selected parameter was the initial spacing of nanopillars, which determines the total number of nanopillars present in the optimization.
Although the nanopillars are allowed to move freely during the optimization process, the initial condition for the structure discussed earlier was a regular grid of nanopillars with a spacing of 300 nm. 
Over a 5~{\textmu}m x 5~{\textmu}m area, this corresponds to a total of 256 nanopillars, or 1280 optimization parameters. 
No nanopillars were added or removed over the course of the optimization in order to maintain a fixed number of optimization parameters. 

In principle, the optimization procedure could be expanded upon to allow for the addition or removal of nanopillars between optimization phases by utilizing the gradients and constraint functions that are already calculated at each iteration.
For example, pillars could be removed if their derivative with respect to both axial lengths are negative but the lengths themselves are already at their lower bound determined by the minimum feature size.
Pillar pairs could be combined if their derivatives indicate they should be closer together but they are limited by the constraint functions described in Eqn. \ref{eq:constraints}. 
And new pillars could be added if the gradient fields, described as the dot product between forward and adjoint electric fields (see Eqn. S1), are large enough to indicate a pillar should exist in an empty region.

In order to determine the effect of nanopillar height and spacing, we performed several additional optimizations varying these pre-selected values. 
The results are quantified in Fig. \ref{fig:comparison}, and the final geometries from each optimization are shown in Figs. S4-S5 in the supplementary information. 
In order to account for variations in convergence, we performed five optimizations for each initial condition in order to quantify the range of potential outcomes. 
The five optimizations used identical starting grids but varied the initial axial lengths of the nanopillars. 
In one optimization, each nanopillar was initialized with a 100-nm-diameter circular cross-section; for the other four, the axial lengths were initially set to random values between 25 nm and 125 nm. 
We did not observe any trend as to whether the uniform or randomized starting axial lengths led to better-performing final structures.

From Fig.~\ref{fig:comparison}a, we observe that the metasurface performance tends to improve with increasing nanopillar height (the pitch was fixed at 300~nm). 
The simulations for 750 nm and 1~{\textmu}m tall nanopillars showed similar average performance; we chose 750-nm-tall nanopillars for subsequent optimizations due to their higher maximum performance and their shorter height being easier to fabricate accurately.

Using in each case 750-nm-tall pillars, Fig. \ref{fig:comparison}b shows the performance as a function of initial pillar pitch.
We observe comparable average performance with an initial nanopillar pitch of 250, 300, and 350 nm, but significantly worse performance at 200 nm and 400 nm. 
The 400 nm pitch has the fewest total nanopillars in the optimization region (144), and the performance was likely poor here due insufficient degrees of freedom. 
Conversely, for an initial pitch of 200~nm, the nanopillar density was likely too high to allow room for the nanopillar positions or sizes to change without violating the constraint functions. 
Amongst the 250 nm, 300 nm, and 350 nm batches, there is a noticeable trend that a larger number of pillars (and therefore more optimization parameters) leads to a larger deviation in performance as a result of different initial conditions, albeit with similar average performance.

Overall, we find that the choice of initial pillar pitch does not strongly impact the final device performance as long as the total number of parameters is not too small (as with the 400 nm batch) and the dynamic range of values for each parameter is not limited by constraints (as with the 200 nm batch).
Additionally, the choice of a 300 nm pitch and 750 nm height we chose based on this analysis is very similar to the choice of a 300 nm pitch and 1{~\textmu}m height chosen to achieve a full $2\pi$ phase response to design the phase-gradient metasurface in Ref. \cite{Huang2019AEmitters}, with the 750 nm height performing only marginally better in the best-case scenario.
We expect that in general, choosing fixed parameters similar to those used for a conventionally-designed device should lead to good results.

\begin{figure}
    \centering
    \includegraphics{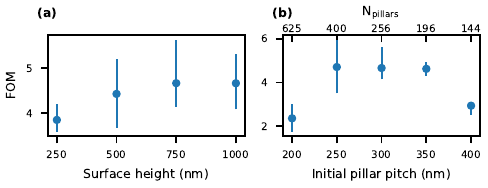}
    \caption{Figure of merit of structures resulting from optimizations performed with different \textbf{(a)} nanopillar heights and \textbf{(b)} initial nanopillar spacings. Each data point corresponds to the results of a batch of five different optimizations with different initial conditions on the nanopillar axial lengths, with error bars extending to the best- and worst-performing structures. The figure of merit calculated in each optimization was the total dipole transmission efficiency into a free-space NA of 0.2, optimized for only one dipole orientation. In \textbf{(a)}, all of the nanopillars had an initial spacing of 300 nm, and in \textbf{(b)}, the nanopillars were 750 nm tall.}
    \label{fig:comparison}
\end{figure}

\subsection{Tolerance to Fabrication Errors}

\begin{figure}
    \centering
    \includegraphics{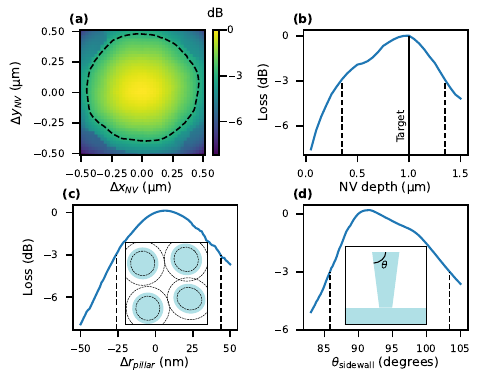}
    \caption{Simulated tolerance of structure performance to the NV-center alignment and fabrication errors, expressed in dB loss from the design target. 3 dB points and contours are marked with dashed lines. \textbf{(a)} Tolerance to in-plane displacements, $\Delta x_{NV}$ and $\Delta y_{NV}$, of the NV center at the correct depth. \textbf{(b)} Loss as a function of the depth of the NV center beneath the bottom of the metasurface. The designed depth of 1~{\textmu}m is shown with a solid line. \textbf{(c)} Loss as a function of a uniform radius perturbation applied simultaneously to every nanopillar in the array, modeling an over- and under-etch of the structure. The inset shows a central 600nm $\times$ 600nm region of the metasurface depicted in Fig.~\ref{fig:concept}b with dashed lines showing pillar sizes at 3 dB loss points. {(d)} Loss as a function of nanopillar sidewall angle as defined in the inset diagram.}
    \label{fig:tolerance}
\end{figure}

We simulated the effects of potential fabrication errors and plotted the results in Fig. \ref{fig:tolerance}. 
The performance losses are expressed in units of decibels (dB) compared to the performance as designed. 
Figures \ref{fig:tolerance}a and \ref{fig:tolerance}b show the performance loss due to lateral and vertical misalignment, respectively. 
The loss is calculated by performing many separate FDTD simulations with the emitter placed at varying positions. 
For lateral misalignments, the 3 dB loss contour occurs for displacements around 470 nm. 
The structure shows similar sensitivity to displacements in the emitter depth.
With the design targeted to a depth of 1~{\textmu}m beneath the metasurface, we predict a loss within 3 dB with the emitter instead placed as shallow as 0.35~{\textmu}m or as deep as 1.35~{\textmu}m.
Lithographic fabrication methods can be used to align the metasurface over a chosen NV center well within these fabrication tolerances\,---\,previous work has successfully aligned solid immersion lenses onto pre-selected NV centers of a particular depth with lateral accuracy less than 100 nm and higher accuracy is possible \cite{Jamali2014MicroscopicMilling}.
If desired, the alignment sensitivity could be further decreased by explicitly incorporating it into the optimization process by co-optimizing the performance from multiple emitter locations. 

We also simulate the effect of common errors in the etching process. 
Fig. \ref{fig:tolerance}c accounts for the effects of over- or under-etching by universally perturbing the axial lengths on each nanopillar in the metastructure. 
Interestingly, the structure actually performs slightly better with slightly larger (under-etched) nanopillars than those given by the final optimization, with peak performance around $\Delta r = +6$ nm. 
This likely occurs due to the constraints placed on nanopillar spacing described in Eqn. \ref{eq:constraints}. 
In the optimized structure, 35 pairs of nanopillars are within 1 nm of the minimum separation, and the optimization would likely have converged on a solution with smaller spacings if the constraints were not enforced. 
In any case, the 3 dB point occurs at perturbed axial lengths $\Delta r$ of +44 nm and -26 nm for under- and over-etched nanopillars, respectively.

Finally, in Fig. \ref{fig:tolerance}d, we simulate the effect on performance if the nanopillar sidewalls are not etched with perfectly vertical, 90-degree angles. Similarly to the over/under-etch analysis, the maximum performance occurs for a sidewall angle of 92.0 degrees, corresponding to nanopillars that are effectively somewhat larger than designed. 
The 3 dB points for sidewall angles are at 85.8 and 103.4 degrees.

\section{Discussion and Outlook}

The goal of the optimized structure is to maximize photon collection from a 1-\textmu m-deep NV center directly into an optical fiber, obviating the need for high-NA free-space optics.
We are not aware of any superior solution for this task.
We simulate an overall transmission efficiency of 3.61\% (including all emitted broadband photons assuming a (100) diamond surface) into the collection aperture of a multi-mode fiber. 
This is a factor of 3.5 improvement over a fiber-coupled SIL or a factor of 26 improvement over a fiber-coupled flat air-diamond interface; a SIL can only reach its peak efficiency when combined with a high-NA optic and remains the best-performing option in that case.
To predict the device performance in an experimental setting, we compare to Ref. \cite{Huang2019AEmitters}, in which the authors measured a saturation photon count rate of 87.3 $\pm$ 2.8 kcps (kilocounts per second) from an NV center imaged through a similar metasurface coupled to a multi-mode fiber.
In simulations, the optimized metasurface outperforms that structure by roughly 60\%, and we would expect to obtain a photon count rate around 150 kcps under otherwise identical experimental conditions and with similar losses from fabrication errors.
Additionally, the metasurface achieves this greater performance while being much more compact (5 \textmu m vs 28 \textmu m diameter) and having a shorter depth of focus (1 \textmu m vs 20 \textmu m), which allows for easier alignment and fabrication.

The many-body shape optimization methodology we present achieves comparable performance to the more common adjoint-based technique of topology optimization.
While topology optimization enforces its fabrication constraints by subsequently applying filters over multiple optimization phases, MBSO allows for constraint functions to be defined explicitly from the parameterization and applied throughout a single-phase optimization.
As a result, the shape-optimized structure converges much faster, using less than half as many sets of simulations as a topology-optimized structure (200 vs 471)\,---\,the computational requirements of both types of adjoint optimization are primarily limited by the number of simulations required.
The shape-optimized structure also required an order of magnitude fewer parameters than the topology-optimized structure (1280 vs 40,401); although this did not strongly impact the computational performance in this case, it could result in significantly faster gradient calculations for larger-scale metasurfaces.
However, while topology optimization requires very little knowledge or assumptions about what a final device might look like, MBSO does require an extra degree of judiciousness in choosing certain reasonable fixed parameters (in this case pillar pitch and height) as discussed in Section \ref{hyperparameters}. 

As an adjoint-based optimization strategy, MBSO's scalability to larger geometrical areas is primarily limited by the computational requirements in memory and the time to perform the full-wave simulations at each iteration.
Several techniques may be used to increase the viability of large-scale optimizations.
Computational advancements such as GPU acceleration can enable larger-scale full-wave simulations \cite{Warren2019ASoftware}.
Past authors have performed shape optimizations on structures an order of magnitude larger than the present work by applying symmetrical boundary conditions and using rectangular elements aligned to the FDTD mesh \cite{Mansouree2021Large-ScaleOptimization} or by using Maxwell solvers based on boundary integral methods \cite{Sideris2019UltrafastMethods, Garza2022FastMethods}\,---\,both strategies are particularly well-suited to shape optimizations.
Other works have avoided full-wave simulations by using local periodic approximations \cite{Pestourie2018InverseMetasurfaces} and coupled mode theory \cite{Zhou2021InverseOptimization} in limited regimes or deep learning methods to predict forward and inverse problems \cite{Zhelyeznyakov2021DeepMetasurfaces, Kanmaz2023Deep-learning-enabledMetasurfaces, Jiang2019GlobalNetwork}.
The MBSO parameterization method could easily be adapted into any of these alternative optimization strategies.

The MBSO methodology is readily applicable to the inverse design a variety of structures, especially when there is a need for specific shapes or distinct elements.
Broadly, it can be used in lieu of topology optimization for any inverse design problem.
In particular, MBSO is naturally suited for the inverse design of photonic structures such as photonic crystals, large-area metasurfaces, and diffraction gratings that are typically composed of arrays of simple shapes.
Additionally, MBSO could be useful for designing reconfigurable metasurfaces\,---\,which have potential applications including imaging, beam steering for LiDAR, and displays\,---\, such that the design space is limited to a fixed number of meta-atoms that can be independently addressed by some tuning mechanism \cite{Gu2022ReconfigurableSuccess}.
For certain materials, it may be challenging to fabricate arbitrary shapes due to facet-selective etching or growth.
In these instances, MBSO could be easily adapted to keep any edges aligned to particular crystal facets in order to optimize structures exactly how they would be fabricated, potentially enabling the inverse design of photonic structures onto new material platforms.

\begin{acknowledgements}
The authors thank Brian Edwards for fruitful conversations and Mathieu Ouellet for assistance with the three-dimensional image in Fig. 1(a).
A.R.K. acknowledges support from the Fontaine Society and National Science Foundation Graduate Research Fellowship Program. 
A.R.K. and L.C.B. acknowledge support from the National Science Foundation under awards ECCS-1842655 and ECCS-2129183.
N.E. acknowledges partial support from the Air Force Office of Scientific Research Multidisciplinary University Research Initiative under grant FA9550-21-1-0312.
\end{acknowledgements}

\bibliography{references, references_manual}

\end{document}

\title{Supplementary Info: Designing metasurface optical interfaces for solid-state qubits using many-body adjoint shape optimization}

\author{Amelia R. Klein}
\affiliation{Department of Electrical and Systems Engineering, University of Pennsylvania, Philadelphia PA 19104, USA}

\author{Nader Engheta}
\affiliation{Department of Electrical and Systems Engineering, University of Pennsylvania, Philadelphia PA 19104, USA}

\author{Lee C. Bassett}
\email{lbassett@seas.upenn.edu}
\affiliation{Department of Electrical and Systems Engineering, University of Pennsylvania, Philadelphia PA 19104, USA}

\maketitle
\onecolumngrid

\section{Gradient calculation for elliptical nanopillars}

In order to calculate the gradients given a forward and adjoint simulation, we integrate a product of the two fields around the boundary of the ellipse as defined using a level set formulation for each ellipse in Equation (5.40) in \cite{Miller2012}, reproduced here in Eqn. \ref{eqn:Millerthesis} where $F$ is the (per-wavelength) figure of merit, $\epsilon_1$ and $\epsilon_2$ are the permittivities inside and outside each nanopillar, $\phi_i$ is a level set describing the ith pillar, and $\mathbf{p_i} = (x_0^{(i)}, y_0^{(i)}, a_i, b_i, \theta_i)$ are the corresponding optimization parameters defining the same pillar. 

\begin{equation}\label{eqn:Millerthesis}
\delta F = -2\mathrm{Re}\sum_i\int\delta \mathbf{p_i}\cdot \frac{\partial\phi_i}{\partial\mathbf{p_i}}\frac{1}{|\nabla\phi_i|}\big[(\epsilon_2 - \epsilon_1)\mathbf{E}_\parallel(\mathbf{x})\cdot\mathbf{E_\parallel^A}(\mathbf{x}) + \big(\frac{1}{\epsilon_1} - \frac{1}{\epsilon_2}\big)\mathbf{D_\perp}(\mathbf{x})\cdot\mathbf{D_\perp^A}(\mathbf{x})\big]dA
\end{equation}

The gradient is calculated by integrating this function over the ellipse boundary, with the forward ($\mathbf{E}_\parallel, \mathbf{D}_\perp$) and adjoint ($\mathbf{E_\parallel^A}, \mathbf{D_\perp^A})$ fields separated into parallel and perpendicular components to the ellipse normal to ensure boundary continuity. To perform the integration, the fields are interpolated using trilinear interpolation from the mesh grid onto points $(x,y)$ given by an ellipse parameterization $u \in [0, 2\pi]$ in Eqn \ref{eqn:ellipseparam} and a normal vector given by \ref{eqn:ellipsenorm}. The integration is performed over the entire 3D surface, with the $z$ components being more trivial as $z \in [0,750\mathrm{nm}]$ for our 750 nm tall nanopillars.

\begin{equation}\label{eqn:ellipseparam}
(x,y) = (x_0^{(i)} + a_i\cos{u}\cos{\theta_i} - b_i\sin u\sin\theta_i, y_0^{(i)} + a_i\cos u\sin\theta_i + b_i\sin u\cos\theta_i)
\end{equation}

\begin{equation}\label{eqn:ellipsenorm}
\hat{n} = <b_i\cos u\cos\theta_i - a_i\sin u\sin\theta_i, a_i\sin u\cos\theta_i + b_i\cos u\sin\theta_i, 0>
\end{equation}

The same fields are integrated over each nanopillar surface for each of the five optimization parameters defining it. The five gradients are distinguished by the additional scaling term $\frac{\partial\phi_i}{\partial\mathbf{p_i}}$ in Eqn. \ref{eqn:Millerthesis}. In order to calculate this term, each ellipse is defined by a level set $\phi_i$ given in Eqn. \ref{eqn:levelset} with $\phi_i = 0$ defining the ellipse boundary.

\begin{subequations}\label{eqn:levelset}
\begin{equation}
\phi_i = \Big(\frac{A_i}{a_i}\Big)^2 + \Big(\frac{B_i}{b_i}\Big)^2 - 1
\end{equation}
\begin{equation}
A_i \equiv (x-x_0^{(i)})\cos\theta_i + (y-y_0^{i})\sin\theta_i
\end{equation}
\begin{equation}
B_i \equiv -(x-x_0^{(i)})\sin\theta_i + (y-y_0^{i})\cos\theta_i
\end{equation}
\end{subequations}

The scaling parameters are then determined by the partial derivatives with respect to the level set formulation, as given in Eqn. \ref{eqn:derivs}.

\begin{subequations}\label{eqn:derivs}
\begin{equation}
\frac{\partial\phi_i}{\partial x_0^{(i)}} = \frac{2A_i}{a^2}\frac{\partial A_i}{\partial x_0^{(i)}} + \frac{2B_i}{b_i^2}\frac{\partial B_i}{\partial x_0^{(i)}} = -\frac{2A_i\cos\theta_i}{a_i^2} + \frac{2B_i\sin\theta_i}{b_i^2}
\end{equation}

\begin{equation}
\frac{\partial\phi_i}{\partial y_0^{(i)}} = \frac{2A_i}{a_i^2}\frac{\partial A_i}{\partial y_0^{(i)}} + \frac{2B_i}{b_i^2}\frac{\partial B_i}{\partial y_0^{(i)}} = -\frac{2A_i\sin\theta_i}{a_i^2} - \frac{2B_i\cos\theta_i}{b_i^2}
\end{equation}

\begin{equation}
\frac{\partial\phi_i}{\partial a_i} = -\frac{2A_i^2}{a_i^3}
\end{equation}

\begin{equation}
\frac{\partial\phi_i}{\partial b_i} = -\frac{2B_i^2}{b_i^3}
\end{equation}

\begin{equation}
\frac{\partial\phi_i}{\partial\theta_i} = \frac{2A_i}{a_i^2}\frac{\partial A_i}{\partial\theta_i} + \frac{2B_i}{b_i^2}\frac{\partial B_i}{\partial \theta_i} = 2A_iB_i\Big(\frac{1}{a_i^2} - \frac{1}{b_i^2}\Big)
\end{equation}
\end{subequations}

\section{Construction of Adjoint Source}

The inverse design strategy allows the calculation of an arbitrary figure of merit and its gradient using one forward simulation and one adjoint simulation. The source for the forward simulation is the normal source for the problem (in this case, an optical dipole emitter), and the source of the adjoint simulation is derived from reciprocity relations and depends on the form of the figure of merit function. From the formulation in \cite{Miller2012}, an adjoint source can be calculated for a figure of merit of the form:

\begin{equation}
    F(\textbf{E}, \textbf{H}) = \int_{\chi}f(\textbf{E}(x), \textbf{H}(x))d^3x
\end{equation}

For such a figure of merit, the adjoint sources should consist of electric sources $\mathbf{P(x)} = \frac{\partial f}{\partial \mathbf{E}}$ and magnetic sources $\mathbf{M(x)} = -\frac{1}{\mu_0}\frac{\partial f}{\partial \mathbf{H}}$ placed over the same measurement region $\chi$. For a figure of merit based on transmission through a surface with incident fields $\mathbf{E_i}, \mathbf{H_i}$, we have:

\begin{equation}\label{eq:transmission_supplement}
F = \iint_{R}\frac{1}{2}\operatorname{Re}\big(\mathbf{E_i}\times\mathbf{H_i^*})\cdot \boldsymbol{\hat{\textbf{z}}}dxdy
\end{equation}

For this, we find source terms as the following, where $\textbf{n}$ is the normal vector to the measured collection plane:

\begin{subequations}
    \begin{equation}
        \textbf{P}(\textbf{x}) = \frac{1}{2}\textbf{H}_i^* \times \textbf{n}
    \end{equation}
    \begin{equation}
        \textbf{M}(\textbf{x}) = \frac{1}{2\mu_0}\textbf{E}_i^* \times \textbf{n}
    \end{equation}
\end{subequations}

Equivalently, we define our sources $\textbf{E}_s, \textbf{H}_s$ in Lumerical FDTD \cite{Lumerical} to be

\begin{subequations}\label{eq:adjointsource}
    \begin{equation}
        \textbf{E}_s  = \frac{1}{2\epsilon}\textbf{H}_{i}^* \times \textbf{n}
    \end{equation}
    \begin{equation}
        \textbf{H}_s = \frac{1}{2\mu_0}\textbf{E}_i^* \times \textbf{n}
    \end{equation}
\end{subequations}

Essentially, the adjoint source consists of the same wave incident on the collection surface reflected backward, to a scaling factor. Importantly, the source terms are only defined within the integration region $R$ and are zero elsewhere. For example, $R$ can be defined as a circle with a diffraction-limited radius to define a focusing efficiency problem. In this case, the adjoint source would consist of the reflected fields incident on that focal region, but zero elsewhere. 

For our figure of merit used in the main text, we instead limit our region $R$ in k-space rather than physical space. For this, the process is identical. The fields are Fourier transformed into k-space, only the components within the desired collection angle are kept, and the fields are inverse Fourier transformed back into physical space to define the adjoint source as described in Eqn. \ref{eq:adjointsource}.

Rigorously, the adjoint source profile is frequency-dependent when optimizing over many frequencies as we do in our work. Such a frequency-dependent field source can be imported directly into Lumerical FDTD when the frequencies are sampled onto a chebyshev grid. 

Alternatively, a figure of merit can be defined as the overlap into a particular mode, as has been done in prior adjoint optimizations in Lumerical FDTD with waveguide modes \cite{Lalau-Keraly2013}, in which case the adjoint source is, to a scaling factor, the same mode sent backwards. The same principles hold when using an arbitrary desired field profile, such as a Gaussian beam, is defined as the ``mode'' in this case. 

We have implemented figure of merit functions and their corresponding adjoint sources in Code 1, Ref. \cite{Lumopt-mbso}. We have done so for the case of the total integrated Poynting vector through a surface, as in Eqn. \ref{eq:transmission_supplement}, and for an overlap with an arbitrarily field profile. Additionally, we have a third figure of merit implementation that performs a sort of hybrid functionality. It uses as its figure of merit for the optimization the total integrated Poytning vector, but uses as the adjoint source a Gaussian beam (or other field profile). Rigorously, this is only an approximate solution, but it allows for the adjoint source to always be well-defined and avoids computational complexity and inaccuracies from using a frequency-dependent source. We find that using a Gaussian approximation to the adjoint source typically yields similar results to the rigorously defined source, but offers faster computation time.

\section{Comparison to Unit Cell Design}

\begin{figure}[h]
\centering
\includegraphics{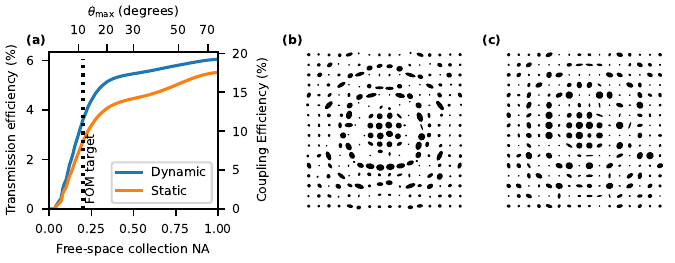}
\caption{Comparison of optimization results in which pillar positions are free parameters (dynamic) versus results in which pillars are fixed in place (static). \textbf{(a)} Plot of transmission efficiency as a function of collection NA for both designs. \textbf{(b)} Final geometry for the dynamic case, as shown previously in the main text. \textbf{(c)} Final geometry of the static case.}
\label{fig:unitcell}
\end{figure}

Our methodology breaks from conventional unit-cell design approaches by allowing the positions of each pillar to be free parameters in addition to their shapes and orientations.
This increases the potential design space and allows for improved results.
In order to quantify this improvement, we ran another set of optimizations in which the pillars are locked in their initial positions.
The results are illustrated in Fig. \ref{fig:unitcell} and compared to the optimized structure in the main text.
The case in which pillars are allowed to remove results in a directional transmission efficiency of 3.61\%, a 41\% increase over the efficiency of 2.56\% demonstrated in the unit-cell design.

\section{Comparison to Topology Optimization}

\begin{figure}[h]
\centering
\includegraphics{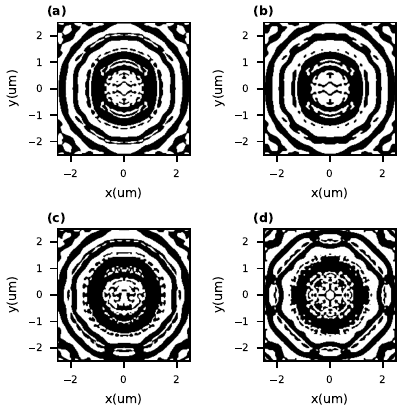}
\caption{Geometries resulting from four different topology optimizations. Regions in black indicate the presence of diamond, while white regions contain air. 50 nm smoothness filters are applied to each during the optimization to remove sharp features. Each structure is a cross-section extruded to a height of 750 nm, analogous to the shape optimizations. In \textbf{(a)}, the structure is optimized to collect from a single dipole oriented in the x-direction. In \textbf{(b)}, the same optimization is performed with the addition of a penalty function enforcing a minimum feature size of 50 nm. In \textbf{(c)}, the structure is optimized for a pair of [100]-oriented dipoles, and in \textbf{(d)}, the structure is optimized for a pair of [111]-oriented dipoles.}
\label{fig:topogeoms}
\end{figure}

We performed a set of several optimizations in order to compare the performance of our many-body shape optimization technique to the more common method of topology optimization, with the results shown in Fig. \ref{fig:topogeoms}. To perform these simulations, we use the topology optimization shipped with the LumOpt add-on to Lumerical FDTD \cite{Lalau-Keraly2013}. Our topology-optimized surfaces are the same height as the nanopillars in our MBSO. Binarization filters are automatically applied over the course of the optimization to result in a structure composed of only diamond and air. Fabrication constraints are not as directly defined using topology optimizations. We set a smoothing filter with a size of 50 nm over the course of all of our optimizations to limit the size of features present. Additionally, we also set a penalization parameter for violations to the minimum feature size; implementing this stricter restriction caused the performance to drop by about half, and the resulting geometry without and with this penalization parameter can be seen in Fig. \ref{fig:topogeoms}a and Fig. \ref{fig:topogeoms}b, with the latter having noticeably fewer small features.

We compare the performance of our topology and shape-optimized structures in Fig. \ref{fig:topocomp}. We find that the performance around the targeted figure of merit at a collection numerical aperture of 0.2 is almost identical, but the topology-optimized structures transmit a higher percentage of light at larger numerical apertures. Both sets of structures applied a minimum feature size of 50 nm. Our choice to use MBSO gave us direct control over the fabrication constraints and the types of output shapes, and this enhanced control did not ultimately come at the cost of performance.

Our MBSO optimization is performed in one continuous gradient-descent optimization run. Topology optimization, contrarily, requires multiple optimization phases in which binarization and smoothing parameters are applied in order to converge on a realisable solution. Because these additional optimization phases were not necessary, we converged on a solution much quicker. Our MBSO-based structure highlighted in the main text and in Fig. \ref{fig:topocomp}a converged using 199 sets of forward and adjoint simulation calls, while the TO-based structure used 471 sets. Additionally, fabrication constraints were manifest at every iteration in the MBSO, whereas the TO structure only became realizable towards the end of the process.

Our MBSO optimization also used far fewer parameters than the topology optimization. For our MBSO, we used an initial 16x16 grid of pillars represented by five parameters each, resulting in a total of 1280 parameters. For our topology optimization, we divided the 5~{\textmu}m $\times$ 5~{\textmu}m region into 25 nm pixels, resulting in a total of 40,401 optimization parameters representing the relative permittivity of each pixel.

\section{Comparison to Solid Immersion Lens}

\begin{figure}[h]
\centering
\includegraphics{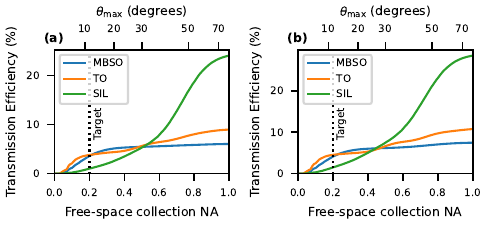}
\caption{Performance comparison of many-body shape optimization (MBSO), topology optimizations (TO), and a solid immersion lens (SIL). In \textbf{(a)}, performance is compared for structures optimized and characterized for dipoles in [100] diamond. In \textbf{(b)}, performance is compared for structures optimized and characterized for dipoles in [111]-oriented diamond.}
\label{fig:topocomp}
\end{figure}

The prototypical structure used for enhancing collection from deep NV centers is the solid immersion lens (SIL). This is simply a hemisphere etched with a NV center at its center, such that any emitted photons are incident normal to the surface and not trapped by total internal reflection. These structures are very effective at transmitting light out of the diamond, but they do \textit{not} modify the light's directionality, and much of it remains at larger angles. We simulated an ideal SIL with centered around an analogous dipole, and compared the results to our optimized structures in Fig. \ref{fig:topocomp}. We find that our structure performs much better at the targeted collection NA of 0.2 (3.6\% vs 1.0\% of total light collected directionally), but the SIL transmits significantly more total light when including large-angle emission (6\% vs 24\% of total light collected). Other works have created topology-optimized structures that have similar performance as the SIL by optimizing for total transmission rather than directional transmission \cite{Wambold2020, Chakravarthi2020Inverse-designedQubits}.

\section{Optimization batches}

\begin{figure}[h!]
\centering
\includegraphics{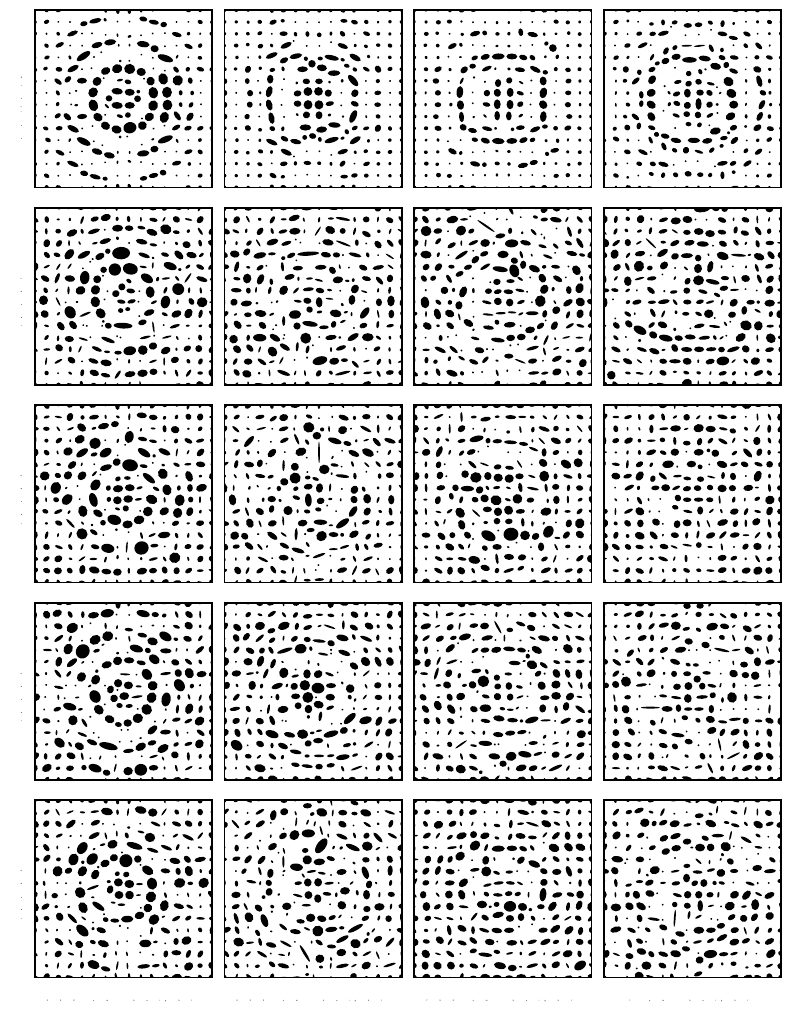}
\caption{Final geometries from batch of optimizations used to generate plot of figure of merit vs pillar height in the main text. Each was optimized to maximize collimated transmission efficiency from a single x-oriented dipole. The five optimizations at each pillar height vary by the initial axial lengths of the pillars: one used uniform axial lengths, and four used randomized axial lengths. The randomized axial lengths are different for each batch in the set.}
\label{fig:geom_vs_height}
\end{figure}

\begin{figure}[h!]
\centering
\includegraphics{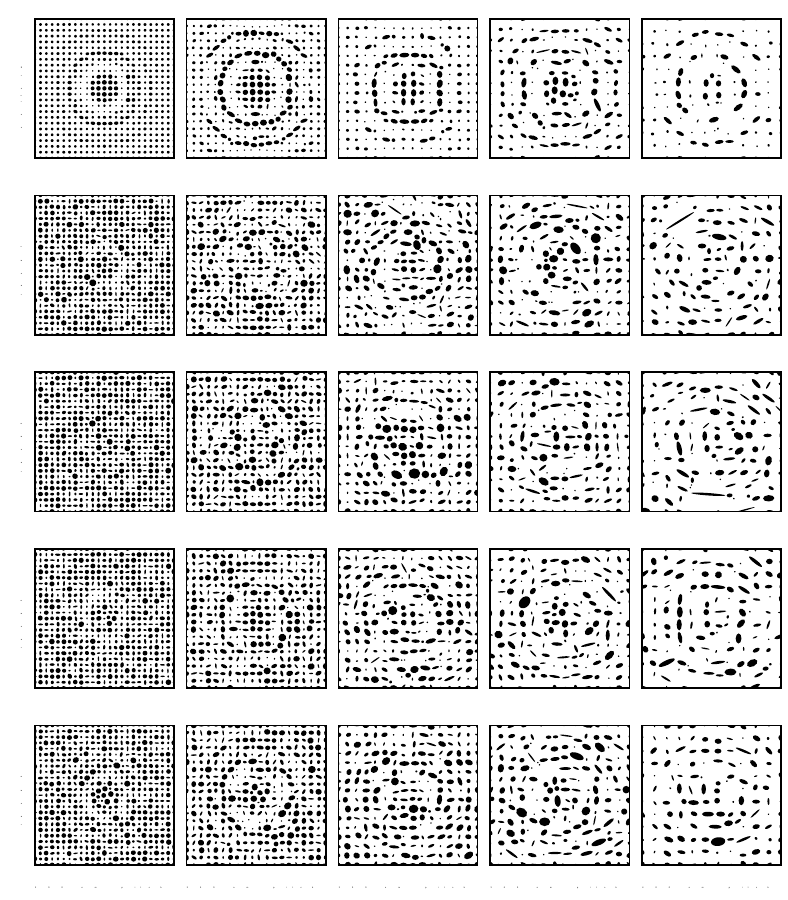}
\caption{Final geometries from batch of optimizations used to generate plot of figure of merit vs pillar pitch in the main text. Each was optimized to maximize collimated transmission efficiency from a single x-oriented dipole. The five optimizations at each pillar pitch vary by the initial axial lengths of the pillars: one used uniform axial lengths, and four used randomized axial lengths. The randomized axial lengths are different for each batch in the set.}
\label{fig:geom_vs_pitch}
\end{figure}

In Fig. 4 in the main text, we compare the results of several batches of optimizations varying the pillar heights and initial pillar pitches. For each chosen height and each chosen pitch, five optimizations were performed with the initial axial lengths of each pillar varied in order to get quantify the range of potential outcomes with the chosen parameters. The resulting geometry from every one of this optimizations is shown here. Fig. \ref{fig:geom_vs_height} shows all resulting geometries for pillar heights of 250, 500, 750, and 1000 nm and Fig. \ref{fig:geom_vs_pitch} shows all resulting geometries from initial pillar spacings of 200 nm, 250 nm, 300 nm, 350 nm, and 400 nm.

\section{NV axis dependence}

\begin{figure}
\centering
\includegraphics{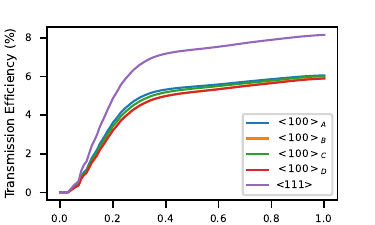}
\caption{Performance of structure in main text under different assumptions of NV axis orientation. Assuming a (100) top surface to the diamond, there are four possible NV axes that give unique sets of dipole orientations, with $<100>_A$ being the choice optimized over. Additionally plotted is the result for a (111) diamond surface with the NV axis oriented directly upwards.}
\label{fig:perorientaiton}
\end{figure}

In the main text, we make the assumption that our diamond has a (100) top face, which is typical of CVD diamond used in NV experiments. The NV center itself is then always along one of the <111> axis orientations, with the possible optical dipole orientations existing in the same (111) plane. We chose one particular NV axis to optimize our structure for, but in practice there are eight potential NV axes orientations for a (100)-oriented diamond: $[111], [\bar{111}], [11\bar{1}], [\bar{11}1], [1\bar{1}1], [\bar{1}1\bar{1}], [\bar{1}11], $and$ [1\bar{11}]$. Only four of these eight uniquely define a plane of potential optical dipole orientations, with the optical dipole moment at room temperature being a mostly random orientation within its allowed plane. These four possible orientations can be distinguished experimentally for a particular NV center by observing its polarization and far-field properties. For simplicity, we can label the four possible sets of optical dipoles for a (100) diamond as $(100)_A, (100)_B, (100)_C, (100)_D$. With $(100)_A$ corresponding to the choice of dipole orientations that we used to optimize our structure in the main text, we plot the performance of all possible sets of orientations in Fig. \ref{fig:perorientaiton}. Additionally, we show the performance if instead the diamond had a (111) top face and the NV axis were aligned vertically. As expected, the performance is highest with the case of the $(111)$ diamond and the orientation we optimized for, $(100)_A$ performs better than the other possible orientations with a $(100)$ diamond. However, the asymmetry of the structure is not strong, and strong transmission is still observed in the other possible $(100)$ orientations. In order to explicitly optimize for any possible dipole orientation with multiple possible axes directions, the optimization could instead be repeated simultaneously optimizing over three, rather than two, orthogonal orientations ($\hat{x}, \hat{y}, \hat{z}$).

\section{Lateral alignment tolerance}

In Fig. 5a of the main text, we characterize the performance of our structure as a function of lateral displacements of the NV center from the center of the metasurface. In order to characterize this performance, we performed a set of simulations with the dipole source placed at each plotted location. In order to calculate the performance, we use a modified version of our figure of merit function in the main text, particularly with respect to the integration region in k-space. From the main text, we have this integration region defined by:

\begin{equation}
R = \{k_x, k_y: \sqrt{k_x^2 + k_y^2}/k_0 \leq \mathrm{NA_{target}}\}
\end{equation}

\begin{figure}
\centering
\includegraphics{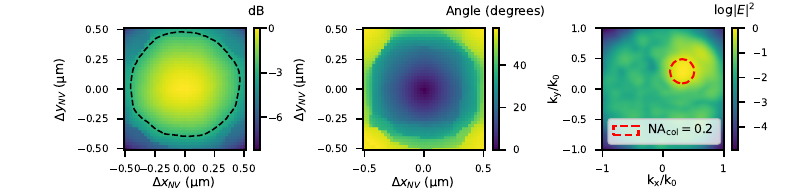}
\caption{\textbf{(a)} Performance of metasurface with respect to lateral NV displacements, copied from Fig. 5a in the main text. \textbf{(b)} Angle from the z-axis of output light as a function of dipole displacement, calculated from simulated field profiles. \textbf{(c)} k-space field profile resulting from an NV center placed at $(x,y) = (-250 \mathrm{nm}, -250 \mathrm{nm})$. The dotted region indicates a NA=0.2 region centered around the peak intensity. This region is noticeably displaced from the origin, indicating light exits the metasurface at an angle from the normal.}
\label{fig:lat_tol}
\end{figure}

When the dipole is displaced laterally, the emission out of the metasurface is no longer oriented perpendicularly, but at some angle. To account for this, we modify our region $R$:

\begin{equation}
R = \{k_x, k_y: \sqrt{(k_x - k_{x0})^2 + (k_y - k_{y0})^2}/k_0 \leq \mathrm{NA_{target}}\}
\end{equation}

where $k_{x0}$ and $k_{y0}$ represent the point in k-space with the peak field power. From $k_{x0}$ and $k_{y0}$, we can calculate the angle from the z-axis of output emission; an optical fiber would need to be aligned matching this same angle to achieve the specified tolerances. For a fiber aligned exactly vertically, there would be much greater performance drops due to lateral NV displacements. We plot the tolerance, output angle, and an example k-space field profile in Fig. \ref{fig:lat_tol}.

\section{Computational Complexity of Constraints}

Unlike topology optimization and simpler shape optimizations, our parameterization utilizes a number of inequality constraint functions $c_{ij}$ to represent the separation between nanopillars, with the number of functions scaling with the number of pillars. As analytical functions, evaluating these constraints takes negligible computation time when compared to the full-wave simulations, but a large number of constraint functions calculated simultaneously can consume a significant amount of memory. For ease of implementation, we applied a constraint function between \textit{every} pair of nanopillars, resulting in a constraint matrix that scales in size as $\mathcal{O}(N^2)$ for $N$ nanopillars and a constraint Jacobian matrix scaling in size as $\mathcal{O}(N^3)$. This poor scaling did not cause issues for our structure but could result in prohibitively large memory usage for larger optimization regions. A dramatically more efficient implementation should instead apply constraint functions only between nearby pairs and use a sparse matrix representation of the constraint Jacobian. the memory usage and computational complexity of carefully implemented constraints should then remain negligible when compared to the full-wave Maxwell solver when scaling up to a larger number of elements.

\bibliography{references, references_manual}